\newcommand\sovast{\ref@jnl{Soviet~Ast.}} 

\RequirePackage[normalem]{ulem} 
\RequirePackage{color}\definecolor{RED}{rgb}{1,0,0}\definecolor{BLUE}{rgb}{0,0,1} 

\documentclass[usenatbib]{mn2e}
\usepackage{url,ulem,times,graphicx,amsmath,amsfonts,amssymb,color,epsfig,epstopdf}
\usepackage{graphics}
\usepackage{epsf}
\usepackage{bm}
\usepackage{color}
\usepackage{rotating}
\usepackage[T1]{fontenc}
\usepackage{ae,aecompl}
 \usepackage{array,multirow}
\usepackage[percent]{overpic}

\usepackage{natbib}
\def\aap{A\&A}

\def\apjs{ApJS}
\def\apjl{ApJL}

\def\lsim{\mathrel{\lower0.6ex\hbox{$\buildrel {\textstyle <}
 \over {\scriptstyle \sim}$}}}
\def\gsim{\mathrel{\lower0.6ex\hbox{$\buildrel {\textstyle >}
 \over {\scriptstyle \sim}$}}}

\def\eone{${\bf e}_{1}$}
\def\etwo{${\bf e}_{2}$}
\def\ethree{${\bf e}_{3}$}

\begin{document}

\title[Planes of satellite galaxies and the cosmic web ]{Planes of satellite galaxies and the cosmic web}
\author[Libeskind et al]
{Noam I Libeskind$^{1}$\thanks{email: {\tt nlibeskind@aip.de}},Yehuda Hoffman$^{2}$, R. Brent Tully$^{3}$, Helene M Courtois$^{4}$,\newauthor Daniel Pomar\`{e}de$^{5}$, Stefan Gottl\"{o}ber$^{1}$, Matthias Steinmetz$^{1}$\\
$^1$Leibniz-Institut f\"ur Astrophysik Potsdam (AIP), An der Sternwarte 16, D-14482 Potsdam, Germany\\
 $^2$Racah Institute of Physics, Hebrew University, Jerusalem 91904, Israel\\
 $^3$Institute for Astronomy (IFA), University of Hawaii, 2680 Woodlawn Drive, HI 96822, US\\
 $^4$University of Lyon; UCB Lyon 1/CNRS/IN2P3; IPN Lyon, France.\\
 $^5$Institut de Recherche sur les Lois Fondamentales de l'Univers, CEA/Saclay, 91191 Gif-sur-Yvette, France.
  }

\date{Accepted  9 June 2015 --- Received 19 March 2015  \\ \newline {\large $\star$ Contact: {\tt nlibeskind@aip.de}}}
\maketitle

 \begin{abstract}
Recent observational studies have demonstrated that the majority of satellite galaxies tend to orbit their hosts on highly flattened, vast, possibly co-rotating planes. Two nearly parallel planes of satellites have been confirmed around the M31 galaxy and around the Centaurus A galaxy, while the Milky Way also sports a plane of satellites. It has been argued that such an alignment of satellites on vast planes is unexpected in the standard ($\Lambda$CDM) model of cosmology if not even in contradiction to its generic predictions. Guided by $\Lambda$CDM numerical simulations, which suggest that satellites are channeled towards hosts along the axis of the slowest collapse as dictated by the ambient velocity shear tensor, we re-examine the planes of local satellites systems within the framework of the local shear tensor derived from the Cosmicflows-2 dataset. The analysis reveals that the Local Group and Centaurus A reside in a filament stretched by the Virgo cluster and compressed by the expansion of the Local Void. Four out of five thin planes of satellite galaxies are indeed closely aligned with the axis of compression induced by the Local Void. Being the less massive system, the moderate misalignment of the Milky Way's satellite plane can likely be ascribed to its greater susceptibility to tidal torques, as suggested by numerical simulations. The alignment of satellite systems in the local universe with the ambient shear field is thus in general agreement with predictions of the $\Lambda$CDM model.\\

\end{abstract}

\section{Introduction}
\label{section:intro}

	The Copernican principle dictates that the universe, at large, is homogeneous and isotropic. Yet the environment on galactic scales is anisotropic since galaxies reside within an complex network of filaments and sheets often dubbed ``The Cosmic Web'' \citep{1996Natur.380..603B}. This anisotropic nature of the galaxy distribution in the Milky Way's (MW) vicinity has long been evident since \cite{1976RGOB..182..241K} and \cite{1976MNRAS.174..695L,LyndenBell1982} noted that the brightest satellites of the MW are distributed in a plane that is roughly perpendicular to the MW disc \citep[although there are difficulties in assessing this planar structure due to obscuration from the Galactic disc, e.g.][]{2004MNRAS.353..639W}. The highly flattened plane of low-luminosity dwarf satellites seen around the MW was considered unique until these magnitudes could be probed elsewhere. \cite{2009Natur.461...66M} recently compiled a photometric and spectroscopic survey known as Pan-Andromeda Archaeological Survey (PAndAS) of the stellar halo of M31, our nearest large galaxy, and uncovered a rich satellite system not unlike our own \citep[see also][]{2011ApJ...740...69C,2012ApJ...758...11C}. These observations revealed that roughly half (13 out of 27) of M31's satellite galaxies belong to a ``vast, thin, co-rotating'' plane \citep{2013Natur.493...62I,2013ApJ...766..120C} viewed nearly edge on from the Earth. \cite{2013Natur.493...31T} further suggested that a trio of dwarfs beyond M31's virial radius (IC1613, IC10 and LGS 3) also aligned on the same plane found by Ibata et al. Follow up analysis by \citet{2013MNRAS.436.2096S} of the remaining 14 satellites of M31 revealed that at least eight of them were also located on a second plane (although not necessarily a co-rotating one). The two planes of satellites around M31 are roughly parallel to each other, offset by around 15 degrees. 
	
	A bit further afield, Centaurus A (hereafter CenA), a massive galaxy with around 30  dwarf satellites, shows a very similar set up to M31: \cite{2015ApJ...802L..25T} found that all but 2 out of 29 of the CenA satellites, for which distances are known, appear to cluster onto two roughly parallel planes (offset by 8 degrees) each incredibly thin (r.m.s. values of $\sim$50kpc), fairly long (extending $\sim 250$kpc) and separated by around 300 kpc. Beyond these three systems, satellite galaxies have been found to exhibit flattened distributions around M81 \citep{2013AJ....146..126C} and NGC 3109 \citep{2013A&A...559L..11B}. Using stacking techniques, similar anisotropic distributions of satellites have been found in large surveys \citep{1997ApJ...478L..53Z} such as the 2dF \citep{2004MNRAS.348.1236S} and SDSS \citep{2005ApJ...628L.101B,2006MNRAS.369.1293Y,2010ApJ...709.1321A}. 
	
Although not identically defined, the evidence for the ubiquitous anisotropic distribution of satellite galaxies has by now a strong observational foundation. The abundance of evidence pointing towards the anisotropic distribution of satellites around large central host galaxies calls for a comprehensive and robust explanation, within the framework of the cold dark matter with dark energy ($\Lambda$CDM) model. Accordingly gravitational instability drives the formation of structure via the amplification of small primordial fluctuations that prevailed in the infant universe \citep{1970A&A.....5...84Z}. In doing so, small initial deviations from perfect spherical symmetry are also amplified. The cosmic web is a manifestation of the anisotropic nature of the instability.

$\Lambda$CDM makes very strong predictions on how the large-scale structure evolves in general, and fully accounts for the collapse of dark matter halos. However its constraints on the distribution of baryons, in particular of stars and stellar systems, are considerably less conclusive. On small scales, this is exemplified by the so-called ``missing satellite'' problem \citep{1999ApJ...522...82K,1999ApJ...524L..19M} and the ``too big to fail'' problem \cite{Boylan11}. A number of solutions to both of these issues have been postulated, many of which rely on correctly approximating the complex interplay between baryons and dark matter \citep[e.g.][]{DiCintio11,2014arXiv1412.2748S}. It is by now clear, that baryons must play a crucial role on these scales.

Accordingly, gas-dynamical numerical simulations \citep[e.g.][]{2015MNRAS.446..521S} and $N$-body runs with semi-analytical models \citep[e.g.][]{2005Natur.435..629S} have achieved unprecedented success in describing the galaxy population as a whole. However such models have continued to struggle to explain the origin of satellite galaxy setups such as those seen in the local universe \citep[although have not altogether failed, see][]{2005A&A...437..383K,2005ApJ...629..219Z,2005MNRAS.363..146L,2007MNRAS.374...16L,2011MNRAS.415.2607D}. Since most $\Lambda$CDM simulations have consistently found it difficult to explain these satellite planes \citep{2012MNRAS.424...80P,2014MNRAS.442.2362P}, a number of authors have suggested alternative formation mechanisms, namely that they are of tidal origin \citep[e.g.][]{2005A&A...431..517K,2011A&A...532A.118P,2007MNRAS.376..387M}, possibly engendered by ancient merger with M31 \citep{2012MNRAS.427.1769F}.

A less controversial issue is how and to what extent the cosmic web imprints its preferred directions on galaxies in general \citep[e.g.][]{2004MNRAS.352..376A}. Recently, \cite{2014MNRAS.443.1274L} showed that satellites are channeled towards their hosts along directions dictated by the large scale structure, irrespective of environment, redshift, or host/subhalo mass. Specifically the eigenvectors of the velocity shear tensor, used also in this work, determine to a large extent, the orientation of the preferred axes along which satellites are accreted \citep[see also][]{2011MNRAS.411.1525L}. 

Motivated by these studies which link the large and small scales of structure formation, we aim to examine the relationship between planes of satellites and the velocity field in the local universe. The distribution of the satellites around host galaxies in general, and around the MW, M31 and CenA in particular, raises one main issues: can such thin partially co-rotating setups be accommodated within the $\Lambda$CDM model or are these anisotropic distributions at odds with generic predictions? Is the {\it observed} large-scale structure velocity field responsible for the peculiar satellite alignments seen around the MW, M31 and Cen A?

\section{Method}

Numerous schemes for characterizing the galaxy distribution on the Megaparsec scale have been suggested, each with their own pros and cons depending on the task at hand. In this work, we describe the cosmic web using peculiar velocities (the dynamical contribution due to the general expansion is not considered). In order to do so we appeal to the Cosmicflows-2 \citep[CF2,][]{2013AJ....146...86T} survey of peculiar velocities. Presently this is the richest and deepest full sky survey of its kind. It enables the reconstruction of the underlying density and 3D velocity fields out to distance exceeding 150 Mpc. The quality of the data allows for the mapping of the nearby universe in great detail: in particular it has established the existence of the Laniakea Supercluster that includes the Local Group and the entire Local Supercluster \citep{2014Natur.513...71T}.

\begin{table}
 \centering
 \begin{tabular}{llrrrc}
\hline
&& SGX &SGY &SGZ&Ref.  \\
\hline
\hline 
 \parbox[t]{2mm}{\multirow{9}{*}{\rotatebox[origin=c]{90}{Local directions}}} \\

&$\hat{r}_{\rm LV}$&-0.205&-0.433&0.866&[1]\\
&$\hat{r}_{\rm Virgo}$&-0.208&0.978&-0.0356&[2]\\
&$\hat{n}_{\rm M31P1}$&-0.339&-0.234&0.912&[3]\\
&$\hat{n}_{\rm M31P2}$&-0.108&-0.411&0.905&''\\
&$\hat{n}_{\rm CAP1}$&-0.135&-0.442&0.886&[4]\\
&$\hat{n}_{\rm CAP2}$&0.079&0.323&-0.943&''\\
&$\hat{n}_{\rm MWP}$&0.532&-0.306&-0.789&[5]\\  
\\

  \hline
 \parbox[t]{2mm}{\multirow{8}{*}{\rotatebox[origin=c]{90}{Shear Tensor}}}  \\
&\eone~(LG) &0.3316&	0.3183&	-0.8881&This Work\\
&\etwo~(LG)& 0.7887&	0.4229&	0.4462& ''\\
&\ethree~(LG)&-0.5175&	0.8484&	0.1108&''\\
&\eone~(CenA)&	0.0834&	0.3205&	-0.9435&''\\  
&\etwo~(CenA)&	0.8087&	0.5315&	0.2521&''\\
&\ethree~(CenA)&	-0.5823&	0.7840&	0.2148&''\\
\\
  \hline
  
 \end{tabular}
 \caption{Directions in the Local Volume in supergalactic coordinates. The upper rows include, $\hat{r}_{\rm LV}$  and $\hat{r}_{\rm Virgo}$: the unit vectors in the directions of the Local Void and the Virgo cluster. $\hat{n}_{\rm M31P1}$ and  $\hat{n}_{\rm M31P2}$: unit normals to the two planes of satellites around M31. $\hat{n}_{\rm CAP1}$ and $\hat{n}_{\rm CAP2}$:  unit normals to the two planes around CenA. $\hat{n}_{\rm MWP}$: unit normal to the MW plane.  The bottom rows include each of the three components of the orthonormal eigenvectors (\eone, \etwo, and \ethree) of the shear field are evaluated at the Local Group and at the location of Cen A. Note that the eigenvectors are non-directional lines and the +/- direction is thus arbitrary.   References for the directions in the top rows are: [1] \citet{2008ApJ...676..184T}; [2] NED database http://ned.ipac.caltech.edu/; [3] \citealt{2013MNRAS.436.2096S}, Table 2; [4] \citealt{Tullysubmitted2015}; [5] \citealt{2013MNRAS.435.2116P}. }
\label{tab:eigens}
\end{table}

\subsection{Wiener Filter reconstruction}
\label{sec:WF}
The 3D velocity field and its associated shear tensor are calculated from the Cosmicflows-2 dataset by means of the Wiener Filter (WF) and constrained realizations \citep[CRs][]{1999ApJ...520..413Z,2012ApJ...744...43C,2013AJ....146...69C}. The $\Lambda$CDM model with WMAP5 \citep{Komatsu09} cosmological parameters (i.e. $\Omega_{0}=0.228$, $\Omega_{b}=0.046$, $\Omega_{\Lambda}=1-\Omega_{0}$,  $H_{0}=100h=70$ km/s/Mpc, $\sigma_{8}=0.812$) is used as the prior model for the reconstruction. The reconstructed WF density and 3D gravitational velocity fields are evaluated in a box of 320 Mpc/$h$ side-length. A Fast Fourier Transform (FFT) is used to construct the CRs, hence periodic boundary conditions are assumed. The construction of the CRs has been done with a box of 640 Mpc/$h$ side-length, to suppress periodic boundary conditions artifacts. Both WF and CRs fields are spanned on a grid of 256$^{3}$. The WF provides the mean field given the data and the prior model (the CF2 dataset and the $\Lambda$CDM/WMAP5 model in our case). 

In order to estimate the robustness (or accuracy) of the WF reconstruction, an ensemble of 20 CRs is then used to sample the scatter around the mean field. In particular it is used to evaluate the error-bars of the eigenvalues (ie not their magnitudes) and scatter in direction of the eigenvectors of the shear tensor. We note that the spatial resolution of the CRs is double that of the WF, and therefore the uncertainties derived from the CRs are evaluated on the scale of 5.0 Mpc/h, compared with the 2.5 Mpc/h of the WF. As such we do not expect the magnitude of the eigenvalues of the two methods to be in agreement, rather the method is used to get a handle on the error bars.

\subsection{Shear Tensor}
Spatial derivatives of the reconstructed peculiar velocity field are taken and the dimensionless shear tensor is defined as 
$\Sigma_{\alpha\beta}=-\frac{1}{2H_{0}}\bigg(\frac{\partial v_{\alpha}}{\partial r_{\beta}}+\frac{\partial v_{\beta}}{\partial r_{\alpha}}\bigg)
$
where $\alpha$ and $\beta$ are the $x, y$, and $z$ components of the velocity $v$ and position $r$ and $H_0$ is Hubble's constant. A minus sign is added for convention. The tensor's sorted eigenvalues ($\lambda_1>\lambda_2>\lambda_3$) describe the strength of compression (positive values) or expansion (negative values) along the eigenvectors of the shear: \eone, \etwo~and \ethree~\citep[see][]{2012MNRAS.425.2049H,2013MNRAS.428.2489L,2014MNRAS.441.1974L}.The spatial derivatives that are used to build the shear tensor are evaluated by the FFT, hence the formal resolution of the shear is 2.5 Mpc/$h$ for the WF and 5.0 Mpc/$h$ for the CRs. It follows that in the present work the entire Local Group (i.e. the MW and the M31) experience the same velocity shear, nearly identical to the one experienced by the adjacent Cen A (see below). 

The eigenvector directions that we accept in this work are compared with published values for the normals to the five satellite planes found around the MW, M31 and Cen A as well as other objects in the local universe. These are presented in the top section of Table.~\ref{tab:eigens}

\section{Results}

\begin{figure*}
 \includegraphics[width=32pc]{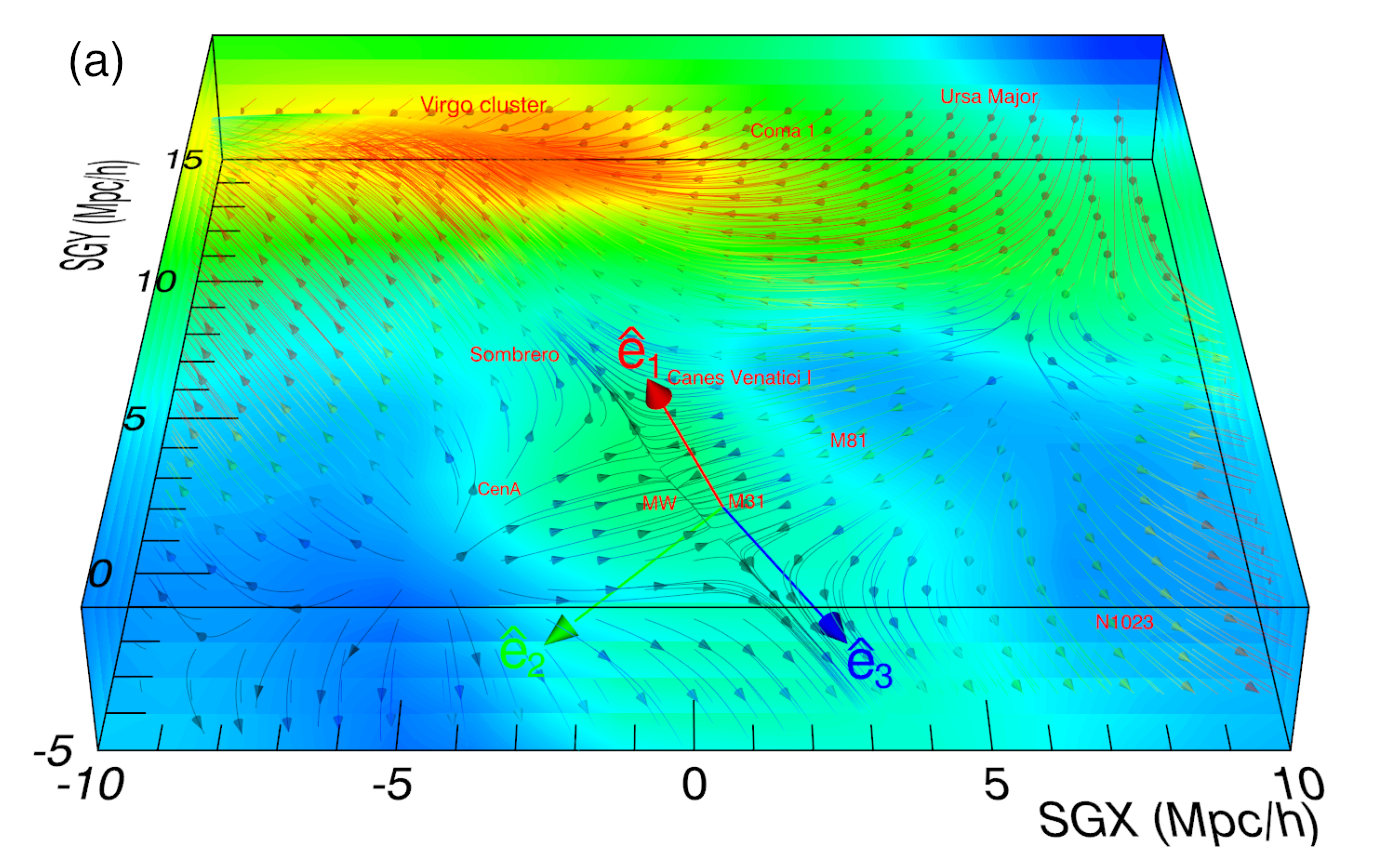}
\includegraphics[width=42pc]{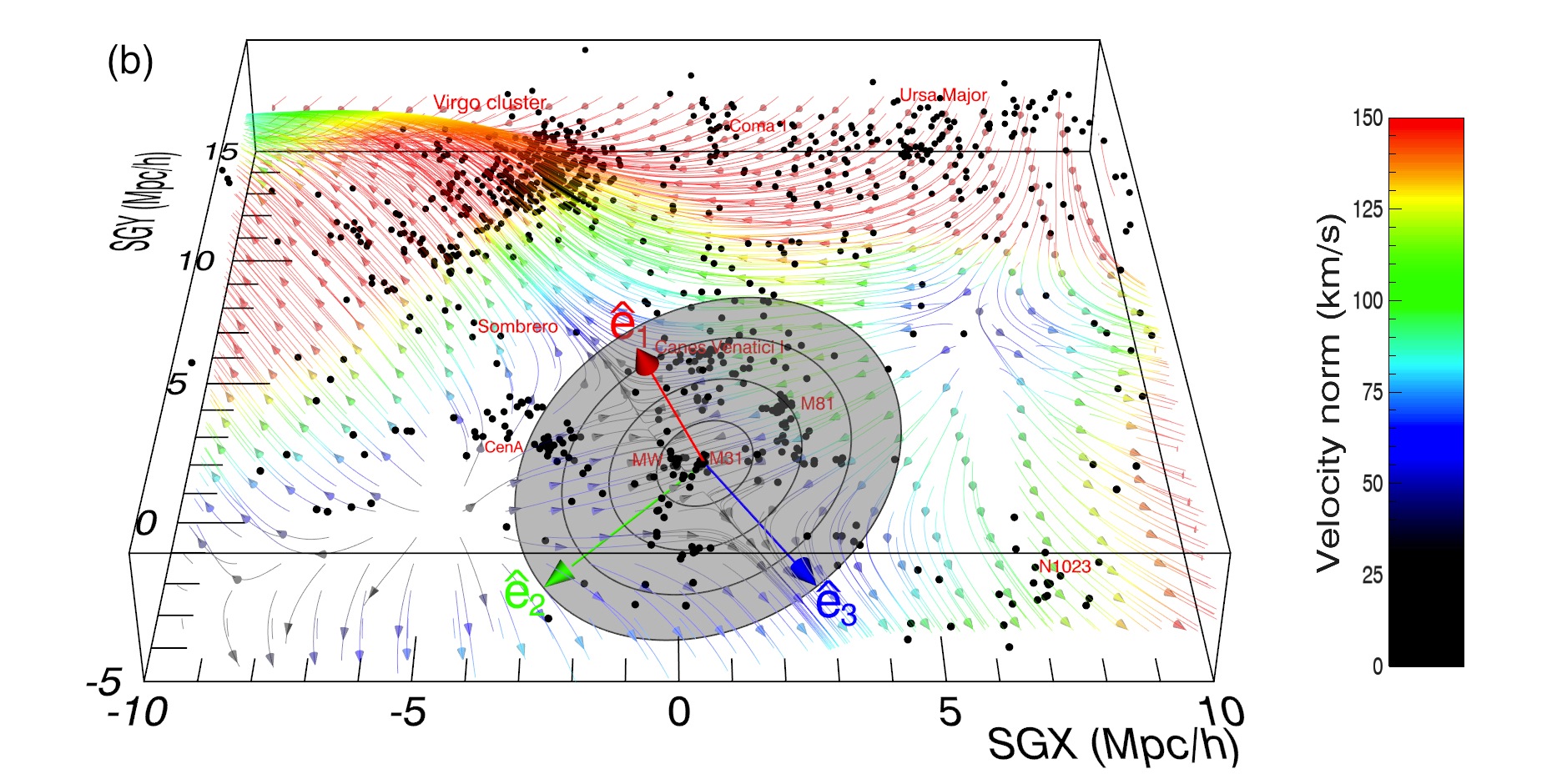}
\caption{The cosmic peculiar velocity and related over-density field in the local universe, reconstructed from the CosmicFlows-2 survey, by means of the Wiener Filter methodology. In both plots stream-lines are colored according to the local speed, shown by the color bar at right. In the top plot (a), the over-density field is colored from the highest density regions (the Virgo Cluster in red) to under-dense ones (voids, in blue). In the bottom plot (b) all galaxies from the V8k survey \citep{2009AJ....138..323T} in this $20\times20\times3$Mpc/$h$ slab are plotted as black dots. The flow towards the Virgo Cluster, as well as the ``local filament'' that bridges the Local Group with the Virgo Cluster, exhibits clear features in both the density and velocity fields. Centaurus A is located on the edge of this filament. The Local Void, which sits above the slab in the positive SGZ direction, pushes down onto the local filament. The shear eigenframe centered on the LG is indicated by the three eigenvectors. The (\etwo-\ethree) plane is shown as the grey disc in (b). The \eone~eigenvector points mostly out of the slab. Note how the local filament and flow toward Virgo are aligned closely with \ethree. In both figures, labels for some well known objects are also included. }
 \label{fig:maps}
 \end{figure*}

A Wiener Filter reconstruction  applied to the CF2 data recovers the underlying 3D gravitationally induced velocity field with an effective resolution of a few Megaparsecs.  At the location of the Local Group and Cen A, the eigenvalues are found to be ($\lambda_1$, $\lambda_2$, $\lambda_3$) = (0.032, 0.018, -0.020) and (0.029, 0.011, -0.025), respectively. Considerable information is encoded in these eigenvalues and their relative magnitudes. The sign, combined with the fact that $\lambda_2$ is much closer to $\lambda_1$ than it is to $\lambda_3$ suggests that the Local Group and Cen A reside in a filament compressed along two eigenvectors (\eone~and \etwo) and expanding along the third (\ethree). This picture is also evident in the spatial cosmography and reconstructed velocity field, shown in Fig.~\ref{fig:maps}. The shear eigenframe at the location of the Local Group is indicated by the three arrows (the \etwo-\ethree~plane is shown by a grey disc).  The shear eigenframe does not change much 3.8 Mpc away at the location of Cen A. It is clear from both the velocity field as well as the density field that both systems reside in a filament, with Cen A being right on the border between a small void and the local filament. The stretching induced by Virgo along \ethree~and the compression due to a void in the direction of \etwo~are clearly manifested. The Local Void is here above the slab, in the positive supergalactic z direction, and its compression is directed along \eone. In the bottom section of Table~\ref{tab:eigens} we present the eigenvectors (\eone, \etwo, \ethree) of the velocity shear tensor experienced by the Local Group and Cen A, in supergalactic coordinates.

\begin{table}
\begin{center}
 \begin{tabular}{lcr}
\hline
property & $|\cos \theta|$ & degrees apart\\
\hline
  
${\bf e}_{3}\cdot\hat{r}_{\rm Virgo}$&0.9330&	$\sim$21.1\\
${\bf e}_{1}\cdot\hat{r}_{\rm Virgo}$&0.2733&	$\sim$74.1\\
${\bf e}_{1}\cdot\hat{r}_{\rm LV}$&0.9898&	$\sim$8.17\\
${\bf e}_{1}\cdot\hat{n}_{\rm M31P1}$&0.9968&	$\sim$4.5\\
${\bf e}_{1}\cdot\hat{n}_{\rm M31P2}$&0.9704&	$\sim$13.9\\
${\bf e}_{1}\cdot\hat{n}_{\rm CAP1}$&0.9879&	$\sim$8.9\\
${\bf e}_{1}\cdot\hat{n}_{\rm CAP2}$&0.9999&	$\sim$0.3\\
${\bf e}_{1}\cdot\hat{n}_{\rm MWP}$&0.7801&	$\sim$38.7\\
\hline
  \hline
 \end{tabular}
 \end{center}
 \caption{Alignment between the shear tensor and relevant directions in the local volume. We examine primarily \eone, the axis of fastest collapse, but also show how the local filament, defined by \ethree, aligns well with the direction of the Virgo Cluster. $\hat{r}_{\rm LV}$  and $\hat{r}_{\rm Virgo}$  are the unit vectors in the directions of the Local Void  and the Virgo cluster. $\hat{n}_{\rm M31P1}$ and  $\hat{n}_{\rm M31P2}$  are the normals to the two planes of satellites around M31. Similarly $\hat{n}_{\rm CAP1}$ and $\hat{n}_{\rm CAP2}$ are the normals to the two planes around CenA. $\hat{n}_{\rm MWP}$ is the normal to the MW satellite plane.}
\label{tab:angles}
\end{table}

\begin{figure*}
 \includegraphics[width=35pc]{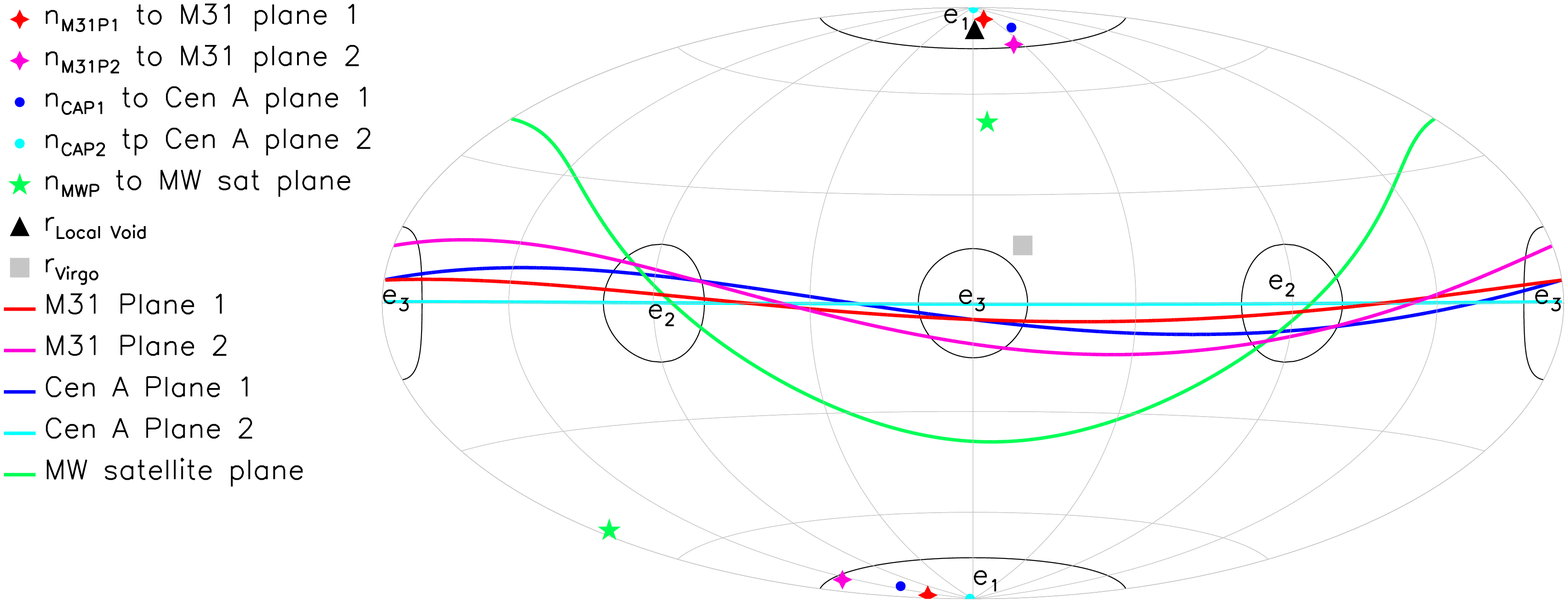}
 \caption{Satellite planes in the local volume projected in the eigenframe of the shear field show a clear tendency to align with the \etwo-\ethree~plane as shown here in an Aitoff-Hammer projection of the unit sphere. Colored curves denote the planes themselves with a pair of corresponding symbols denoting the $\pm$ directions of the normals to each plane. The \eone, \etwo, and \ethree~axes are indicated by the labels. Black curves around these labels show regions of 15 degrees. The angles formed between these satellite planes and the shear eigenframe are given in Tab.~\ref{tab:angles}. Four out of five satellite planes in the Local Volume are well aligned with their local shear field. Note that one Centaurus A satellite plane (CAP2, cyan curve) is within less than a degree of the \etwo-\ethree~plane. The Milky Way satellite plane on the other hand, is rotated about the \etwo~axis by around 38 degrees. This offset can be attributed to a smaller Milky Way halo mass. The direction toward the Local Void ($\hat{r}_{\rm Local~Void}$) is also close to \eone, while the direction toward Virgo ($\hat{r}_{\rm Virgo}$), viewed either from Cen A or the Local Group, is close to \ethree.}
 \label{fig:aits}
 \end{figure*}

The shear tensor eigenvectors provide the natural principal frame of reference within which the anisotropies and preferred directions of the environments of the Local Group and Cen A can be studied. How these planes align within the shear eigen-frame, can be quantified by examining the angle made by each plane's normal and the eigenvectors. Fig.~\ref{fig:aits} shows how the five planes of satellites in the local universe (the MW satellite plane, and the two around M31 and Cen A) are oriented with respect to the principal directions defined by the local shear tensor. The relationships are quantified in Table.~\ref{tab:angles} and described here. The direction towards the center of the Local Void is aligned with \eone. The spine of the local filament (\ethree) points close to the direction of the Virgo Cluster. Most interestingly, the local flow dictates the preferred directions of the M31 and Cen A satellites. The normals to two planes around M31 and Cen A lie very close to \eone, all within $\sim15$ degrees (with one of Cen A's planes being within $\sim1$ degree!) The normal to the MW satellite plane is not as well aligned and is offset by around 38 degrees. We provide a possible explanation for this mis-alignment in section~\ref{sec:sims}. The close alignment of four out of five satellite planes of the local universe with the (\etwo-\ethree) plane is clearly visible. It should be added that the line connecting the MW to M31 lies within the (\etwo-\ethree) plane, well away from \eone~(and  $\sim$40 degrees away from \ethree).

The following picture emerges: the Local Group and Cen A reside in a filament defined primarily by the dark matter distribution. This filament is stretched by the Virgo Cluster, while being compressed from the supergalactic pole by the Local Void and from the sides by smaller voids. The four satellite planes that surround the two most massive galaxies, M31 and Cen A, lie within or close to the (\etwo-\ethree) plane. Both of the M31 satellite planes, the two Cen A satellite planes, and the MW-M31 orbital plane are, to within observational uncertainties for the Local Group spin, roughly parallel with each other and with the (\etwo-\ethree) plane. It follows that influences on scales of tens of Megaparsecs, notably the Virgo Cluster and the Local Void leave their imprint on the anisotropies that characterize the satellite distribution around Cen A and M31.

\subsection{Statistical significance in the estimated shear eigenvalues and eigenvectors}
\label{sec:stats}
As mentioned in section~\ref{sec:WF}, we estimate how robust the determination of the WF shear field is by using CRs. In the standard model of cosmology the over-density and velocity fields, on large enough scales, constitute random Gaussian fields. It follows that the individual matrix elements of the shear tensor constitute random Gaussian variables whose mean value is zero. The eigenvalues of the tensor are not Gaussian variables, and therefore the mean value of each eigenvalue of the ensemble of CRs differs from the corresponding eigenvalue calculated from the WF reconstruction. In particular the mean amplitude of the CRs eigenvalues is expected to be larger than for the WF. The WF provides a more conservative estimation of the eigenvalues than that of the CRs and the WF value is used here to characterize the environment of the Local Group. The CRs are used to estimate the statistical uncertainties for the WF, which is presented in section~\ref{sec:stats}

The shear tensor has been evaluated for the 20 CRs. Table~\ref{tab:crs} presents the mean and the standard deviation of the eigenvalues ($\lambda_{1}$ , $\lambda_{2}$ , $\lambda_{3}$), as well as the differences in eigenvalues ($\lambda_{1} -\lambda_{2}$ and $\lambda_{2}-\lambda_{3}$) and the stability of each eigenvector (${\bf e}_{i}^{\rm WF}\cdot{\bf e}_{i}^{\rm CR}$). The filamentary nature of the local environment is thus confirmed at around the 3$\sigma$ level. 
The table further shows the mean and standard deviation of the alignment of each eigenvector of the CRs with the corresponding eigenvector of the WF field. The stability in the direction of each eigenvector is statistically very significant. Note that the direction of the filament, given by \ethree, is statistically the most robust eigenvector.

\begin{table}
\begin{center}
 \begin{tabular}{lr}
\hline
quantity & mean value $\pm$ std. dev.\\
\hline
  $\lambda_{1}$& 	0.148$\pm$0.038\\
$\lambda_{2}$&	0.051$\pm$0.039\\
$\lambda_{3}$&	-0.160$\pm$0.033\\
$\lambda_{1} -\lambda_{2}$&	0.097$\pm$0.042\\
$\lambda_{2} - \lambda_{3}$&	0.211$\pm$0.061\\
${\bf e}_{1}^{\rm WF}\cdot{\bf e}_{1}^{\rm CR}$&0.890$\pm$0.137\\ 
${\bf e}_{2}^{\rm WF}\cdot{\bf e}_{2}^{\rm CR}$&0.892$\pm$0.136\\
${\bf e}_{3}^{\rm WF}\cdot{\bf e}_{3}^{\rm CR}$&0.923$\pm$0.047\\
\hline
  \hline
 \end{tabular}
 \end{center}
 \caption{Stability of the Wiener Filter reconstruction: The mean value $\pm$ the standard deviation of the eigenvalues, the eigenvalue difference and the alignment of WF and CR  eigenvectors. Given the different smoothings in the two methods (CR versus WF) we do not expect the eigenvalues to have the same magnitude rather, its the small value of the standard deviation which is important.}
\label{tab:crs}
\end{table}

\subsection{The case for the MW mis-alignment}
\label{sec:sims}
As shown in Fig.~\ref{fig:aits}, the MW satellite plane is not as well aligned with the shear field as these other four systems. The normal to this plane appears to have been rotated (around \etwo) by about 38 degrees. This offset may not be a surprise if MW is less massive than its partner M31 as suggested by a number of studies \citep{2007MNRAS.379..755S,2008ApJ...684.1143X,2014ApJ...794...59K,2014MNRAS.443.2204P,2014MNRAS.445.3133P,2014A&A...562A..91P} This is because the satellite system of a less massive halo in a pair like the LG, is dynamically less stable and more prone to torques, in this case that act in the \eone-\ethree~plane.

In order to gauge how the alignment of a plane of satellites with the shear eigenframe may depend on the host halo mass, we turn to numerical simulations, specifically the Bolshoi simulation \citep{2011ApJ...740..102K}, a large cosmological box (of periodic side length $L=250$Mpc/$h$) populated with 2048$^3$ dark matter (DM) particles, each with a particle mass of $\sim1.9\times10^{8}M_{\odot}$. The simulation was run with cosmological parameters that are consistent with WMAP5 and the $z=0$ snapshot is considered. After particles are grouped into DM haloes by use of the BDM halo finder\footnote{Much post-processed data from the Bolshoi (and other simulations) including halo and subhalo catalogues, Merger Trees, and velocity and density fields is publicly available: http://www.cosmosim.org/cms/simulations/multidark-project/}  \citep{1997astro.ph.12217K,2013AN....334..691R}, we divide the halos into two sub-samples that mimic pairs of Local Group galaxies and Cen A. 

To identify Local Groups, we search for all pairs of galaxies: (1) with masses between $5\times10^{11}$ and $5\times10^{12}M_{\odot}$; (2) with a separation between 0.5 Mpc and 1.5 Mpc; (3) that have no halo with mass $> 10^{14}M_{\odot}$ within 10 Mpc of the pair barycenter; (4) and that have no other halo with mass $> 10^{11} M_{\odot}$ within 0.5 Mpc of either candidate. To identify Cen A look-a-likes we simply find all haloes (1) whose mass is between $5\times10^{12}M_{\odot}$ and $5\times10^{13}M_{\odot}$ and (2) that have no mass greater than itself within 3 Mpc. These criteria are designed to be fairly general so as not to impose strong a priori constraints on any subsample, yet are designed to match some of the more accepted characteristics of these galaxies.

\begin{figure}
 \includegraphics[width=20pc]{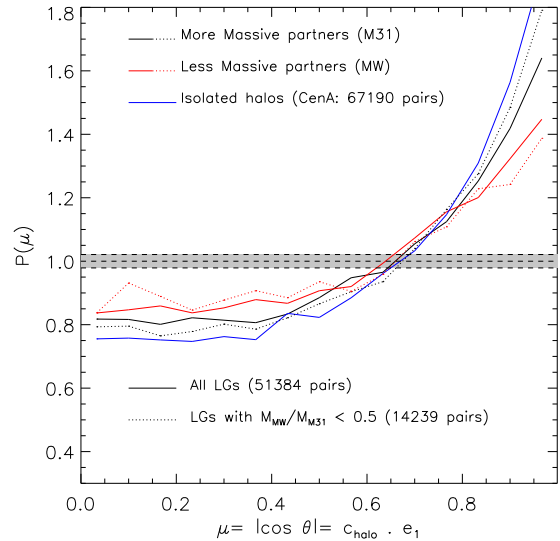}
 \caption{The alignment of dark matter halos with the shear field in a large cosmological simulation shows similar behavior to that observed in the local universe. The probability distribution function of the cosine of the angle formed between the short axis of a halo ($c_{\rm halo}$) and the axis of fastest collapse (\eone) of the surrounding shear field is shown for isolated massive Cen A-like haloes (blue) and for the more massive (black) and less massive (red) partners in LG-like set ups. The dotted line shows the alignment of each LG partner when only LGs with mass ratios less than 1:2 are considered. The shaded gray region represents the scatter about $P(\mu)=1$ that one would expect from a 100,000 uniform distributions of the same size ($\sim$50,000 pairs), thus making these probability distributions highly incompatible with random ones.}
 \label{fig:sims}
\end{figure}

Once the ``Cen A'' and ``Local Group'' samples are amassed, their shapes are computed as defined by the inertia tensor. The shortest axis of the ellipsoid $c_{\rm halo}$ is assumed to point in the same direction as the normal to the plane of satellites. The shear tensor is then computed from the velocity field on a regular grid with a 2.5Mpc smoothing \citep[see previous work,][]{2013MNRAS.428.2489L} and the angle between $c_{\rm halo}$ and the eigenframe is examined. In Fig.~\ref{fig:sims} we present the distribution of the cosine of the angle formed between $c_{\rm halo}$ and \eone. 

Fig.~\ref{fig:sims} shows that the alignment between the less massive Local Group partner (red), the more massive Local Group partner (black) and isolated Cen A-like halos (blue). The dotted curves show the alignment of the two Local Group partners when only pairs with a mass ratio of less than 1:2 are considered \citep[motivated by ][]{2014MNRAS.443.2204P}. A number of interesting points can be inferred here. Isolated massive Cen A-like haloes are well aligned with the shear field - their $c_{\rm halo}$ axes are roughly 25\% more likely to be tilted towards the \eone~axes than expected from a random distribution, making them inconsistent with a uniform distribution at the 6$\sigma$ level. Local Group partners, too, show a tendency to have their $c_{\rm halo}$ axes aligned with the \eone~axis. The more massive Local Group partner (M31-proxy) is around 18\% more likely than a random distribution to be aligned with the shear eigenframe. The less massive partner (MW proxy) has a significantly weaker alignment with the shear eigenframe, with just 14\% better aligned than a random distribution. This result is true for all mass ratios, but is exacerbated for Local Groups where the mass ratio is less then 1:2 (dotted curve). Indeed, with these mass ratios, the more massive partner is around 20\% more aligned than a random distribution; the less massive partner is just 9\% more aligned than a uniform distribution.

The conclusions we arrive at from looking at how DM halo shapes are oriented with respect to the shear field are clear: the more massive the object, the better the alignment; the more isolated the halo, the better the alignment; in Local Group like pairs, the more massive halo is better aligned than the less massive one; when the mass ratio is far from 1:1, the difference in alignment for the more and less massive partner grows. Indeed for mass ratios of 1:10 or more (not shown) the less massive Local Group partner shows no alignment at all.

\section{Summary and conclusions}

Vast planes of dwarf galaxies have been discovered throughout the local universe. First spotted around the MW, \citep{1976RGOB..182..241K,1976MNRAS.174..695L,LyndenBell1982,2013MNRAS.435.2116P}, a pair were found around M31 \citep{2013Natur.493...62I,2013ApJ...766..120C,2013MNRAS.436.2096S} and most recently, a pair was found near Centaurus A (\citealt{2015AJ....149...54T,2015ApJ...802L..25T}). In the case of the MW and at least one of M31's planes, the dwarfs appear to be co-rotating \citep{MetzKroupaLibeskind2008,2014MNRAS.442.2362P}. 

We have applied the Wiener Filter algorithm \citep{1999ApJ...520..413Z} to the CosmicFlows-2 survey of peculiar velocities \citep{2013AJ....146...86T}, to reconstruct the full density and velocity field in the local universe.  Such a reconstruction has allowed us to examine how the five known planes of dwarfs are embedded in the cosmic web. We have found that the main cosmographic feature of the local universe is a filament that stretches from the Virgo Cluster, some 16 Mpc away, across the Local Group that is not seen in the galaxy distribution but clearly evident in the velocity field and dark matter density field. Our main result is that {\it four out of five of the planes of dwarf galaxies that have been discovered are well aligned with the main direction of compression, defined by the principal axis of collapse (\eone) of the shear tensor.} This direction is close to the direction of expansion of the local void. One of the dwarf planes, that of the Milky Way, is not as well aligned with the shear. In Section~\ref{sec:sims} we suggest that this misalignment may be attributed to its lower halo mass, and consequent susceptibility to torques which appear to have rotated the Milky Way satellite plane about the \etwo~axis by around 38 degrees. The main conclusions that we can thus draw, is that structure on the scale of $\sim10$Mpc appears to be having a direct effect on small scale structures that are on the order of the virial radius.

How does such a cosmographic picture fit into the $\Lambda$CDM model? \cite{2014MNRAS.443.1274L} showed that in numerical simulations, the infall direction of sub-halos onto their host halos is determined by the  eigenvectors of the velocity shear tensor (\eone, \etwo, and \ethree). Subhalo accretion displays a clear tendency to occur well away from \eone, the axis of fastest collapse, and preferentially in the (\etwo-\ethree) plane with a propensity to be aligned with \ethree. This accretion pattern extends over all redshifts and over scales much larger than the virial radius of a given halo. Furthermore it happens irrespective of the host or subhalo mass, or environment: namely in voids, sheets, filaments, or clusters (hence it is known as ``universal accretion'') - it is characterized only by the directions of the shear eigenvectors. Alignments of subhaloes with the cosmic web in the Sloan Digital Sky Survey tend to support the idea of universal accretion \citep{2014Natur.511..563I,2015ApJ...799..212L,2015arXiv150202046T}. 

Although weakened by processes associated with virialization, the consequence of universal accretion is also seen in the present epoch distribution of sub-halos within their hosts in numerical simulations such as \cite{2012MNRAS.421L.137L} or \cite{2015arXiv150107764S}. Such flattened distributions as seen in the real universe are not readily found but are to some extent reproducible if the criteria to define them is kept vague and not bound by so-called {\it a posteriori} statistics \citep[e.g.][]{2005MNRAS.363..146L,2005ApJ...629..219Z,2009MNRAS.399..550L,2011MNRAS.415.2607D,2014arXiv1412.2748S,2011MNRAS.413.3013L,2005A&A...437..383K,2013MNRAS.429.1502W,2014arXiv1410.7778C,2015ApJ...800...34G}. A number of these studies, admittedly, find that the criteria needed for a successful $z=0$ match are likely to only produce transient, unstable planes when examined at higher $z$. What is much more stable is the alignment of haloes with the eigenvectors of the cosmic web. Spin and halo shape are directly built by the stresses and strains due to the gravitational deformation of the cosmic velocity field. This is clearly seen in numerical simulations of structure formation and now for the first time demonstrated for the local universe, a region whose velocity field we are able to probe with high accuracy.

That satellites occupy primarily the (\etwo-\ethree) plane can be thought of as a general consequence of the formation of ``Zeldovich pancakes'' \citep{1970A&A.....5...84Z}, wherein matter first collapses into vast cosmic sheets, defined primarily by \eone. On smaller scales, these sheets are composed of small clumps that hierarchically merge according to the ``bottom-up'' scenario, suggested by the $\Lambda$CDM model \citep{1978MNRAS.183..341W}. Since initial collapse has already occurred along \eone, not much material is left to be accreted from this direction. Instead haloes are guided along the (\etwo-\ethree) plane, towards density peaks where their ultimate fate is to merge.

While an examination of the directions in the local universe has led us to an understanding of how satellite planes are aligned, a number of open issues remain. What determines the extreme thinness and/or apparent co-rotation of some of these planes? Are the planes seen in $\Lambda$CDM simulations stable or do they precess and evaporate in time? Also, the azimuthal distribution of M31's satellites is lop-sided, with many galaxies found on the near side, closer to the MW. Are these galaxies reacting the the MW's presence, or does this aspect of the problem have a different origin? We leave these important challenges to future work.

\section*{Acknowledgments}
NIL is supported by the Deutsche Forschungs Gemeinschaft. 
YH has been partially supported by the Israel Science Foundation (1013/12).
HC acknowledge support from the Lyon Institute of Origins under grant ANR-10-LABX-66.
The {\it Bolshoi} simulation was run on the Pleiades supercomputer at the NASA Ames Research Center.

\bibliography{Allrefs}

\begin{thebibliography}{77}
\expandafter\ifx\csname natexlab\endcsname\relax\def\natexlab#1{#1}\fi

\bibitem[{{Agustsson} \& {Brainerd}(2010)}]{2010ApJ...709.1321A}
{Agustsson} I., {Brainerd} T.~G., 2010, \apj, 709, 1321

\bibitem[{{Aubert} {et~al}\mbox{.}(2004){Aubert}, {Pichon}, \&
  {Colombi}}]{2004MNRAS.352..376A}
{Aubert} D., {Pichon} C., {Colombi} S., 2004, \mnras, 352, 376

\bibitem[{{Bellazzini} {et~al}\mbox{.}(2013){Bellazzini}, {Oosterloo},
  {Fraternali}, \& {Beccari}}]{2013A&A...559L..11B}
{Bellazzini} M., {Oosterloo} T., {Fraternali} F., {Beccari} G., 2013, \aap,
  559, L11

\bibitem[{{Bond} {et~al}\mbox{.}(1996){Bond}, {Kofman}, \&
  {Pogosyan}}]{1996Natur.380..603B}
{Bond} J.~R., {Kofman} L., {Pogosyan} D., 1996, \nat, 380, 603

\bibitem[{{Boylan-Kolchin} {et~al}\mbox{.}(2011){Boylan-Kolchin}, {Bullock}, \&
  {Kaplinghat}}]{Boylan11}
{Boylan-Kolchin} M., {Bullock} J.~S., {Kaplinghat} M., 2011, \mnras, 415, L40

\bibitem[{{Brainerd}(2005)}]{2005ApJ...628L.101B}
{Brainerd} T.~G., 2005, \apjl, 628, L101

\bibitem[{{Cautun} {et~al}\mbox{.}(2014){Cautun}, {Wang}, {Frenk}, \&
  {Sawala}}]{2014arXiv1410.7778C}
{Cautun} M., {Wang} W., {Frenk} C.~S., {Sawala} T., 2014, ArXiv e-prints

\bibitem[{{Chiboucas} {et~al}\mbox{.}(2013){Chiboucas}, {Jacobs}, {Tully}, \&
  {Karachentsev}}]{2013AJ....146..126C}
{Chiboucas} K., {Jacobs} B.~A., {Tully} R.~B., {Karachentsev} I.~D., 2013, \aj,
  146, 126

\bibitem[{{Conn} {et~al}\mbox{.}(2012){Conn}, {Ibata}, {Lewis}, {Parker},
  {Zucker}, {Martin}, {McConnachie}, {Irwin}, {Tanvir}, {Fardal}, {Ferguson},
  {Chapman}, \& {Valls-Gabaud}}]{2012ApJ...758...11C}
{Conn} A.~R. {et~al.}, 2012, \apj, 758, 11

\bibitem[{{Conn} {et~al}\mbox{.}(2011){Conn}, {Lewis}, {Ibata}, {Parker},
  {Zucker}, {McConnachie}, {Martin}, {Irwin}, {Tanvir}, {Fardal}, \&
  {Ferguson}}]{2011ApJ...740...69C}
{Conn} A.~R. {et~al.}, 2011, \apj, 740, 69

\bibitem[{{Conn} {et~al}\mbox{.}(2013){Conn}, {Lewis}, {Ibata}, {Parker},
  {Zucker}, {McConnachie}, {Martin}, {Valls-Gabaud}, {Tanvir}, {Irwin},
  {Ferguson}, \& {Chapman}}]{2013ApJ...766..120C}
{Conn} A.~R. {et~al.}, 2013, \apj, 766, 120

\bibitem[{{Courtois} {et~al}\mbox{.}(2012){Courtois}, {Hoffman}, {Tully}, \&
  {Gottl{\"o}ber}}]{2012ApJ...744...43C}
{Courtois} H.~M., {Hoffman} Y., {Tully} R.~B., {Gottl{\"o}ber} S., 2012, \apj,
  744, 43

\bibitem[{{Courtois} {et~al}\mbox{.}(2013){Courtois}, {Pomar{\`e}de}, {Tully},
  {Hoffman}, \& {Courtois}}]{2013AJ....146...69C}
{Courtois} H.~M., {Pomar{\`e}de} D., {Tully} R.~B., {Hoffman} Y., {Courtois}
  D., 2013, \aj, 146, 69

\bibitem[{{Deason} {et~al}\mbox{.}(2011){Deason}, {McCarthy}, {Font}, {Evans},
  {Frenk}, {Belokurov}, {Libeskind}, {Crain}, \&
  {Theuns}}]{2011MNRAS.415.2607D}
{Deason} A.~J. {et~al.}, 2011, \mnras, 415, 2607

\bibitem[{{Di Cintio} {et~al}\mbox{.}(2011){Di Cintio}, {Knebe}, {Libeskind},
  {Yepes}, {Gottl{\"o}ber}, \& {Hoffman}}]{DiCintio11}
{Di Cintio} A., {Knebe} A., {Libeskind} N.~I., {Yepes} G., {Gottl{\"o}ber} S.,
  {Hoffman} Y., 2011, \mnras, 417, L74

\bibitem[{{Fouquet} {et~al}\mbox{.}(2012){Fouquet}, {Hammer}, {Yang}, {Puech},
  \& {Flores}}]{2012MNRAS.427.1769F}
{Fouquet} S., {Hammer} F., {Yang} Y., {Puech} M., {Flores} H., 2012, \mnras,
  427, 1769

\bibitem[{{Gillet} {et~al}\mbox{.}(2015){Gillet}, {Ocvirk}, {Aubert}, {Knebe},
  {Libeskind}, {Yepes}, {Gottl{\"o}ber}, \& {Hoffman}}]{2015ApJ...800...34G}
{Gillet} N., {Ocvirk} P., {Aubert} D., {Knebe} A., {Libeskind} N., {Yepes} G.,
  {Gottl{\"o}ber} S., {Hoffman} Y., 2015, \apj, 800, 34

\bibitem[{{Hoffman} {et~al}\mbox{.}(2012){Hoffman}, {Metuki}, {Yepes},
  {Gottl{\"o}ber}, {Forero-Romero}, {Libeskind}, \&
  {Knebe}}]{2012MNRAS.425.2049H}
{Hoffman} Y., {Metuki} O., {Yepes} G., {Gottl{\"o}ber} S., {Forero-Romero}
  J.~E., {Libeskind} N.~I., {Knebe} A., 2012, \mnras, 425, 2049

\bibitem[{{Ibata} {et~al}\mbox{.}(2014){Ibata}, {Ibata}, {Famaey}, \&
  {Lewis}}]{2014Natur.511..563I}
{Ibata} N.~G., {Ibata} R.~A., {Famaey} B., {Lewis} G.~F., 2014, \nat, 511, 563

\bibitem[{{Ibata} {et~al}\mbox{.}(2013){Ibata}, {Lewis}, {Conn}, {Irwin},
  {McConnachie}, {Chapman}, {Collins}, {Fardal}, {Ferguson}, {Ibata}, {Mackey},
  {Martin}, {Navarro}, {Rich}, {Valls-Gabaud}, \&
  {Widrow}}]{2013Natur.493...62I}
{Ibata} R.~A. {et~al.}, 2013, \nat, 493, 62

\bibitem[{{Kafle} {et~al}\mbox{.}(2014){Kafle}, {Sharma}, {Lewis}, \&
  {Bland-Hawthorn}}]{2014ApJ...794...59K}
{Kafle} P.~R., {Sharma} S., {Lewis} G.~F., {Bland-Hawthorn} J., 2014, \apj,
  794, 59

\bibitem[{{Kang} {et~al}\mbox{.}(2005){Kang}, {Mao}, {Gao}, \&
  {Jing}}]{2005A&A...437..383K}
{Kang} X., {Mao} S., {Gao} L., {Jing} Y.~P., 2005, \aap, 437, 383

\bibitem[{{Klypin} \& {Holtzman}(1997)}]{1997astro.ph.12217K}
{Klypin} A., {Holtzman} J., 1997, ArXiv Astrophysics e-prints

\bibitem[{{Klypin} {et~al}\mbox{.}(1999){Klypin}, {Kravtsov}, {Valenzuela}, \&
  {Prada}}]{1999ApJ...522...82K}
{Klypin} A., {Kravtsov} A.~V., {Valenzuela} O., {Prada} F., 1999, \apj, 522, 82

\bibitem[{{Klypin} {et~al}\mbox{.}(2011){Klypin}, {Trujillo-Gomez}, \&
  {Primack}}]{2011ApJ...740..102K}
{Klypin} A.~A., {Trujillo-Gomez} S., {Primack} J., 2011, \apj, 740, 102

\bibitem[{{Komatsu} {et~al}\mbox{.}(2009){Komatsu}, {Dunkley}, {Nolta},
  {Bennett}, {Gold}, {Hinshaw}, \& {et. al}.}]{Komatsu09}
{Komatsu} E., {Dunkley} J., {Nolta} M.~R., {Bennett} C.~L., {Gold} B.,
  {Hinshaw} G., {et. al}., 2009, \apjs, 180, 330

\bibitem[{{Kroupa} {et~al}\mbox{.}(2005){Kroupa}, {Theis}, \&
  {Boily}}]{2005A&A...431..517K}
{Kroupa} P., {Theis} C., {Boily} C.~M., 2005, \aap, 431, 517

\bibitem[{{Kunkel} \& {Demers}(1976)}]{1976RGOB..182..241K}
{Kunkel} W.~E., {Demers} S., 1976, in Royal Greenwich Observatory Bulletin,
  Vol. 182, The Galaxy and the Local Group, {R.~J.~Dickens, J.~E.~Perry,
  F.~G.~Smith, \& I.~R.~King}, ed., pp. 241--+

\bibitem[{{Lee} \& {Choi}(2015)}]{2015ApJ...799..212L}
{Lee} J., {Choi} Y.-Y., 2015, \apj, 799, 212

\bibitem[{{Libeskind} {et~al}\mbox{.}(2007){Libeskind}, {Cole}, {Frenk},
  {Okamoto}, \& {Jenkins}}]{2007MNRAS.374...16L}
{Libeskind} N.~I., {Cole} S., {Frenk} C.~S., {Okamoto} T., {Jenkins} A., 2007,
  \mnras, 374, 16

\bibitem[{{Libeskind} {et~al}\mbox{.}(2005){Libeskind}, {Frenk}, {Cole},
  {Helly}, {Jenkins}, {Navarro}, \& {Power}}]{2005MNRAS.363..146L}
{Libeskind} N.~I., {Frenk} C.~S., {Cole} S., {Helly} J.~C., {Jenkins} A.,
  {Navarro} J.~F., {Power} C., 2005, \mnras, 363, 146

\bibitem[{{Libeskind} {et~al}\mbox{.}(2009){Libeskind}, {Frenk}, {Cole},
  {Jenkins}, \& {Helly}}]{2009MNRAS.399..550L}
{Libeskind} N.~I., {Frenk} C.~S., {Cole} S., {Jenkins} A., {Helly} J.~C., 2009,
  \mnras, 399, 550

\bibitem[{{Libeskind} {et~al}\mbox{.}(2013){Libeskind}, {Hoffman},
  {Forero-Romero}, {Gottl{\"o}ber}, {Knebe}, {Steinmetz}, \&
  {Klypin}}]{2013MNRAS.428.2489L}
{Libeskind} N.~I., {Hoffman} Y., {Forero-Romero} J., {Gottl{\"o}ber} S.,
  {Knebe} A., {Steinmetz} M., {Klypin} A., 2013, \mnras, 428, 2489

\bibitem[{{Libeskind} {et~al}\mbox{.}(2014{\natexlab{a}}){Libeskind},
  {Hoffman}, \& {Gottl{\"o}ber}}]{2014MNRAS.441.1974L}
{Libeskind} N.~I., {Hoffman} Y., {Gottl{\"o}ber} S., 2014{\natexlab{a}},
  \mnras, 441, 1974

\bibitem[{{Libeskind} {et~al}\mbox{.}(2012){Libeskind}, {Hoffman}, {Knebe},
  {Steinmetz}, {Gottl{\"o}ber}, {Metuki}, \& {Yepes}}]{2012MNRAS.421L.137L}
{Libeskind} N.~I., {Hoffman} Y., {Knebe} A., {Steinmetz} M., {Gottl{\"o}ber}
  S., {Metuki} O., {Yepes} G., 2012, \mnras, 421, L137

\bibitem[{{Libeskind} {et~al}\mbox{.}(2014{\natexlab{b}}){Libeskind}, {Knebe},
  {Hoffman}, \& {Gottl{\"o}ber}}]{2014MNRAS.443.1274L}
{Libeskind} N.~I., {Knebe} A., {Hoffman} Y., {Gottl{\"o}ber} S.,
  2014{\natexlab{b}}, \mnras, 443, 1274

\bibitem[{{Libeskind} {et~al}\mbox{.}(2011){Libeskind}, {Knebe}, {Hoffman},
  {Gottl{\"o}ber}, {Yepes}, \& {Steinmetz}}]{2011MNRAS.411.1525L}
{Libeskind} N.~I., {Knebe} A., {Hoffman} Y., {Gottl{\"o}ber} S., {Yepes} G.,
  {Steinmetz} M., 2011, \mnras, 411, 1525

\bibitem[{{Lovell} {et~al}\mbox{.}(2011){Lovell}, {Eke}, {Frenk}, \&
  {Jenkins}}]{2011MNRAS.413.3013L}
{Lovell} M.~R., {Eke} V.~R., {Frenk} C.~S., {Jenkins} A., 2011, \mnras, 413,
  3013

\bibitem[{{Lynden-Bell}(1976)}]{1976MNRAS.174..695L}
{Lynden-Bell} D., 1976, \mnras, 174, 695

\bibitem[{{Lynden-Bell}(1982)}]{LyndenBell1982}
{Lynden-Bell} D., 1982, The Observatory, 102, 202

\bibitem[{{McConnachie} {et~al}\mbox{.}(2009){McConnachie}, {Irwin}, {Ibata},
  {Dubinski}, {Widrow}, {Martin}, {C{\^o}t{\'e}}, {Dotter}, \& {et
  al}.}]{2009Natur.461...66M}
{McConnachie} A.~W. {et~al.}, 2009, \nat, 461, 66

\bibitem[{{Metz} \& {Kroupa}(2007)}]{2007MNRAS.376..387M}
{Metz} M., {Kroupa} P., 2007, \mnras, 376, 387

\bibitem[{{Metz} {et~al}\mbox{.}(2008){Metz}, {Kroupa}, \&
  {Libeskind}}]{MetzKroupaLibeskind2008}
{Metz} M., {Kroupa} P., {Libeskind} N.~I., 2008, \apj, 680, 287

\bibitem[{{Moore} {et~al}\mbox{.}(1999){Moore}, {Ghigna}, {Governato}, {Lake},
  {Quinn}, {Stadel}, \& {Tozzi}}]{1999ApJ...524L..19M}
{Moore} B., {Ghigna} S., {Governato} F., {Lake} G., {Quinn} T., {Stadel} J.,
  {Tozzi} P., 1999, \apjl, 524, L19

\bibitem[{{Pawlowski} {et~al}\mbox{.}(2014){Pawlowski}, {Famaey}, {Jerjen},
  {Merritt}, {Kroupa}, {Dabringhausen}, {L{\"u}ghausen}, {Forbes}, {Hensler},
  {Hammer}, {Puech}, {Fouquet}, {Flores}, \& {Yang}}]{2014MNRAS.442.2362P}
{Pawlowski} M.~S. {et~al.}, 2014, \mnras, 442, 2362

\bibitem[{{Pawlowski} \& {Kroupa}(2013)}]{2013MNRAS.435.2116P}
{Pawlowski} M.~S., {Kroupa} P., 2013, \mnras, 435, 2116

\bibitem[{{Pawlowski} {et~al}\mbox{.}(2012){Pawlowski}, {Kroupa}, {Angus}, {de
  Boer}, {Famaey}, \& {Hensler}}]{2012MNRAS.424...80P}
{Pawlowski} M.~S., {Kroupa} P., {Angus} G., {de Boer} K.~S., {Famaey} B.,
  {Hensler} G., 2012, \mnras, 424, 80

\bibitem[{{Pawlowski} {et~al}\mbox{.}(2011){Pawlowski}, {Kroupa}, \& {de
  Boer}}]{2011A&A...532A.118P}
{Pawlowski} M.~S., {Kroupa} P., {de Boer} K.~S., 2011, \aap, 532, A118

\bibitem[{{Pe{\~n}arrubia} {et~al}\mbox{.}(2014){Pe{\~n}arrubia}, {Ma},
  {Walker}, \& {McConnachie}}]{2014MNRAS.443.2204P}
{Pe{\~n}arrubia} J., {Ma} Y.-Z., {Walker} M.~G., {McConnachie} A., 2014,
  \mnras, 443, 2204

\bibitem[{{Piffl} {et~al}\mbox{.}(2014{\natexlab{a}}){Piffl}, {Binney},
  {McMillan}, {Steinmetz}, {Helmi}, {Wyse}, {Bienaym{\'e}}, \& {et
  al}}]{2014MNRAS.445.3133P}
{Piffl} T., {Binney} J., {McMillan} P.~J., {Steinmetz} M., {Helmi} A., {Wyse}
  R.~F.~G., {Bienaym{\'e}} O., {et al}, 2014{\natexlab{a}}, \mnras, 445, 3133

\bibitem[{{Piffl} {et~al}\mbox{.}(2014{\natexlab{b}}){Piffl}, {Scannapieco},
  {Binney}, {Steinmetz}, {Scholz}, {Williams}, {de Jong}, \& {et
  al}}]{2014A&A...562A..91P}
{Piffl} T., {Scannapieco} C., {Binney} J., {Steinmetz} M., {Scholz} R.-D.,
  {Williams} M.~E.~K., {de Jong} R.~S., {et al}, 2014{\natexlab{b}}, \aap, 562,
  A91

\bibitem[{{Riebe} {et~al}\mbox{.}(2013){Riebe}, {Partl}, {Enke},
  {Forero-Romero}, {Gottl{\"o}ber}, {Klypin}, {Lemson}, {Prada}, {Primack},
  {Steinmetz}, \& {Turchaninov}}]{2013AN....334..691R}
{Riebe} K. {et~al.}, 2013, Astronomische Nachrichten, 334, 691

\bibitem[{{Sales} \& {Lambas}(2004)}]{2004MNRAS.348.1236S}
{Sales} L., {Lambas} D.~G., 2004, \mnras, 348, 1236

\bibitem[{{Sawala} {et~al}\mbox{.}(2014){Sawala}, {Frenk}, {Fattahi},
  {Navarro}, {Bower}, {Crain}, {Dalla Vecchia}, {Furlong}, {Helly}, {Jenkins},
  {Oman}, {Schaller}, {Schaye}, {Theuns}, {Trayford}, \&
  {White}}]{2014arXiv1412.2748S}
{Sawala} T. {et~al.}, 2014, ArXiv e-prints

\bibitem[{{Schaye} {et~al}\mbox{.}(2015){Schaye}, {Crain}, {Bower}, {Furlong},
  {Schaller}, {Theuns}, {Dalla Vecchia}, \& {et al}}]{2015MNRAS.446..521S}
{Schaye} J., {Crain} R.~A., {Bower} R.~G., {Furlong} M., {Schaller} M.,
  {Theuns} T., {Dalla Vecchia} C., {et al}, 2015, \mnras, 446, 521

\bibitem[{{Shaya} \& {Tully}(2013)}]{2013MNRAS.436.2096S}
{Shaya} E.~J., {Tully} R.~B., 2013, \mnras, 436, 2096

\bibitem[{{Shi} {et~al}\mbox{.}(2015){Shi}, {Wang}, \&
  {Mo}}]{2015arXiv150107764S}
{Shi} J., {Wang} H., {Mo} H., 2015, ArXiv e-prints

\bibitem[{{Smith} {et~al}\mbox{.}(2007){Smith}, {Ruchti}, {Helmi}, {Wyse},
  {Fulbright}, {Freeman}, {Navarro}, {Seabroke}, \& {et
  al.}}]{2007MNRAS.379..755S}
{Smith} M.~C. {et~al.}, 2007, \mnras, 379, 755

\bibitem[{{Springel} {et~al}\mbox{.}(2005){Springel}, {White}, {Jenkins},
  {Frenk}, {Yoshida}, {Gao}, {Navarro}, {Thacker}, {Croton}, {Helly},
  {Peacock}, {Cole}, {Thomas}, {Couchman}, {Evrard}, {Colberg}, \&
  {Pearce}}]{2005Natur.435..629S}
{Springel} V. {et~al.}, 2005, \nat, 435, 629

\bibitem[{{Tempel} {et~al}\mbox{.}(2015){Tempel}, {Guo}, {Kipper}, \&
  {Libeskind}}]{2015arXiv150202046T}
{Tempel} E., {Guo} Q., {Kipper} R., {Libeskind} N.~I., 2015, ArXiv e-prints

\bibitem[{{Tully}(2013)}]{2013Natur.493...31T}
{Tully} R.~B., 2013, \nat, 493, 31

\bibitem[{{Tully}(2015)}]{2015AJ....149...54T}
{Tully} R.~B., 2015, \aj, 149, 54

\bibitem[{{Tully} {et~al}\mbox{.}(2014){Tully}, {Courtois}, {Hoffman}, \&
  {Pomar{\`e}de}}]{2014Natur.513...71T}
{Tully} R.~B., {Courtois} H., {Hoffman} Y., {Pomar{\`e}de} D., 2014, \nat, 513,
  71

\bibitem[{{Tully} {et~al}\mbox{.}(2013){Tully}, {Courtois}, {Dolphin},
  {Fisher}, {H{\'e}raudeau}, {Jacobs}, {Karachentsev}, {Makarov}, {Makarova},
  {Mitronova}, {Rizzi}, {Shaya}, {Sorce}, \& {Wu}}]{2013AJ....146...86T}
{Tully} R.~B. {et~al.}, 2013, \aj, 146, 86

\bibitem[{{Tully} \& {et}(2015)}]{Tullysubmitted2015}
{Tully} R.~B., {et} a., 2015, ArXiv e-prints

\bibitem[{{Tully} {et~al}\mbox{.}(2015){Tully}, {Libeskind}, {Karachentsev},
  {Karachentseva}, {Rizzi}, \& {Shaya}}]{2015ApJ...802L..25T}
{Tully} R.~B., {Libeskind} N.~I., {Karachentsev} I.~D., {Karachentseva} V.~E.,
  {Rizzi} L., {Shaya} E.~J., 2015, \apjl, 802, L25

\bibitem[{{Tully} {et~al}\mbox{.}(2009){Tully}, {Rizzi}, {Shaya}, {Courtois},
  {Makarov}, \& {Jacobs}}]{2009AJ....138..323T}
{Tully} R.~B., {Rizzi} L., {Shaya} E.~J., {Courtois} H.~M., {Makarov} D.~I.,
  {Jacobs} B.~A., 2009, \aj, 138, 323

\bibitem[{{Tully} {et~al}\mbox{.}(2008){Tully}, {Shaya}, {Karachentsev},
  {Courtois}, {Kocevski}, {Rizzi}, \& {Peel}}]{2008ApJ...676..184T}
{Tully} R.~B., {Shaya} E.~J., {Karachentsev} I.~D., {Courtois} H.~M.,
  {Kocevski} D.~D., {Rizzi} L., {Peel} A., 2008, \apj, 676, 184

\bibitem[{{Wang} {et~al}\mbox{.}(2013){Wang}, {Frenk}, \&
  {Cooper}}]{2013MNRAS.429.1502W}
{Wang} J., {Frenk} C.~S., {Cooper} A.~P., 2013, \mnras, 429, 1502

\bibitem[{{White} \& {Rees}(1978)}]{1978MNRAS.183..341W}
{White} S.~D.~M., {Rees} M.~J., 1978, \mnras, 183, 341

\bibitem[{{Willman} {et~al}\mbox{.}(2004){Willman}, {Governato}, {Dalcanton},
  {Reed}, \& {Quinn}}]{2004MNRAS.353..639W}
{Willman} B., {Governato} F., {Dalcanton} J.~J., {Reed} D., {Quinn} T., 2004,
  \mnras, 353, 639

\bibitem[{{Xue} {et~al}\mbox{.}(2008){Xue}, {Rix}, {Zhao}, {Re Fiorentin},
  {Naab}, {Steinmetz}, {van den Bosch}, {Beers}, {Lee}, {Bell}, {Rockosi},
  {Yanny}, {Newberg}, {Wilhelm}, {Kang}, {Smith}, \&
  {Schneider}}]{2008ApJ...684.1143X}
{Xue} X.~X. {et~al.}, 2008, \apj, 684, 1143

\bibitem[{{Yang} {et~al}\mbox{.}(2006){Yang}, {van den Bosch}, {Mo}, {Mao},
  {Kang}, {Weinmann}, {Guo}, \& {Jing}}]{2006MNRAS.369.1293Y}
{Yang} X., {van den Bosch} F.~C., {Mo} H.~J., {Mao} S., {Kang} X., {Weinmann}
  S.~M., {Guo} Y., {Jing} Y.~P., 2006, \mnras, 369, 1293

\bibitem[{{Zaritsky} {et~al}\mbox{.}(1997){Zaritsky}, {Smith}, {Frenk}, \&
  {White}}]{1997ApJ...478L..53Z}
{Zaritsky} D., {Smith} R., {Frenk} C.~S., {White} S.~D.~M., 1997, \apjl, 478,
  L53

\bibitem[{{Zaroubi} {et~al}\mbox{.}(1999){Zaroubi}, {Hoffman}, \&
  {Dekel}}]{1999ApJ...520..413Z}
{Zaroubi} S., {Hoffman} Y., {Dekel} A., 1999, \apj, 520, 413

\bibitem[{{Zel'dovich}(1970)}]{1970A&A.....5...84Z}
{Zel'dovich} Y.~B., 1970, \aap, 5, 84

\bibitem[{{Zentner} {et~al}\mbox{.}(2005){Zentner}, {Kravtsov}, {Gnedin}, \&
  {Klypin}}]{2005ApJ...629..219Z}
{Zentner} A.~R., {Kravtsov} A.~V., {Gnedin} O.~Y., {Klypin} A.~A., 2005, \apj,
  629, 219

\end{thebibliography}
\end{document}